\documentclass[preprint]{elsarticle}

\usepackage{amsmath}
\usepackage{amssymb}
\usepackage{subfigure}
\usepackage[export]{adjustbox}

\journal{Nuclear Physics B}

\newcommand\nn{\nonumber}
\newcommand\eq[1] {\begin{align} #1 \end{align}}

\newcommand{\br}[1]{\left( #1 \right)}
\newcommand{\brs}[1]{\left[ #1 \right]}
\newcommand{\brf}[1]{\left\{ #1 \right\}}
\newcommand{\brm}[1]{\left| #1 \right|}

\newcommand{\bra}[1]{\left< #1 \right|}
\newcommand{\ket}[1]{\left| #1 \right>}

\newcommand{\Sp}{\mbox{Sp}}

\renewcommand{\Re}{\mbox{Re}}
\renewcommand{\Im}{\mbox{Im}}

\newcommand{\vv}[1]{{\bf #1}}

\newcommand{\dd}[1]{{\hat #1}}        

\newcommand{\M} {{\cal M}} 

\newcommand{\GeV}{\mbox{GeV}}
\newcommand{\MeV}{\mbox{MeV}}

\newcommand{\eV} {\mbox{eV}}

\newcommand{\pb}{\mbox{pb}}

\begin{document}

\begin{frontmatter}

\title{The cross section of the process $e^+e^- \to p\bar{p}$ in the vicinity of charmonium $\psi(3770)$ including three-gluon and $D$-meson loop contributions}

\author{Yu.M.~Bystritskiy}
\address{Joint Institute for Nuclear Research, 141980 Dubna, Moscow Region, Russia}
\ead{bystr@theor.jinr.ru}

\begin{abstract}
The total cross section of the process $e^+e^- \to p\bar{p}$ is calculated within the energy range
close to the mass of $\psi(3770)$ charmonium state. It was shown that the main contribution to this cross section
comes from three gluon mechanism which is also responsible of the large phase with respect to the Born amplitude.
This phase provides the characteristic dip behaviour of the cross section in contrast to the usual Breit-Wigner
peak shape. OZI-allowed mechanism with D-mesons in the intermediate state was also estimated and gives
relatively small contribution to the cross section and to the phase.
\end{abstract}


\end{frontmatter}

\section{Introduction}

The electron--positron colliders allow to study different physical processes
with final particles in pure $J^{PC}=1^{--}$ state. For example, single
charmonium production gives the possibility to look at the relativistic
bound state of c-quarks, elaborate and test new ideas of confinement
description and to search for possible exotic admixtures in the wave function
of the state.

One of the intriguing states of charmonium is $\psi(3770)$ which
was studied by many collaborations (for example, by
KEDR-VEPP-4M \cite{Baldin:2008zz, Anashin:2011kq}, CLEO \cite{Ge:2008aa}
and more recently by BES III \cite{Ablikim:2008zz, Ablikim:2014jrz}).
Especially the latter one \cite{Ablikim:2014jrz} made precise measurement
of the total cross section at the specific kinematics of
$\psi(3770)$ mass and showed that instead of a Breit--Wigner peak one can see
a dip in the dependence of the total invariant mass squared $s$ (see Fig.~2 in \cite{Ablikim:2014jrz}).
This observation immediately lead to the conclusion that there must be some
mechanism which generates large relative phase $\phi$ between resonant and continuum
terms in the amplitude.
And indeed soon some possible explanation of this phase was suggested
in Ref.~\cite{Ahmadov:2013ova} where is was attributed to the OZI-violated
three gluon mechanism of charmonium
$\psi(3770)$ transition into final proton--antiproton state.
This type of mechanism was already shown to give large phase
for another charmonium $\chi_2(^3P_2)$ in Ref.~\cite{Kuraev:2013swa}:
it was shown that two gluon mechanism can produce the phase of order of $\phi \sim -90^\circ$.

Here we want to continue the elaboration of three gluon model
from Ref.~\cite{Ahmadov:2013ova} and to refine $D$-meson loop calculation.
We compare our estimations of the total cross section with the precise experimental data
of $\psi(3770)$ production obtained at BES III \cite{Ablikim:2014jrz}.

The paper is organized in the following manner:
in Section~\ref{sec.Born} the total cross section of the process
$e^+e^- \to p\bar{p}$ in Born approximation (that is so called continuum part) is obtained
and some kinematical notations are introduced;
Section~\ref{sec.PsiIntermediateState} is the main calculation part of the paper,
the extra resonant contribution to the amplitude is added and the formalism of how to
take it into account in the cross section is developed.
There are two subsections:
Subsection~\ref{sec.DMesonLoopMechanism} describes the details of calculation
of possible OZI-allowed mechanism with $D$-meson loop,
while Subsection~\ref{sec.ThreeGluonMechanism} briefly reminds some key steps of
three gluon mechanism calculation from \cite{Ahmadov:2013ova}
(with some minor corrections of typos and errors in \cite{Ahmadov:2013ova});
Section~\ref{sec.Numerical} gives some numerical estimations and
comparison of our calculation with experimental data from BES~III \cite{Ablikim:2014jrz};
Section~\ref{sec.Conclusion} summarizes our results and proposes some possible
extension of this work in the future.

\section{Born approximation}
\label{sec.Born}

We consider the process of proton-antiproton pair creation in an
electron-positron collisions:
\eq{
    e^+(q_+)+e^-(q_-) \to p(p_+)+\bar{p}(p_-),
    \label{eq.Process}
}
where quantities in parenthesis are the 4-momenta of the corresponding particles.
\begin{figure}
	\centering
    \subfigure[]{\includegraphics[width=0.4\textwidth]{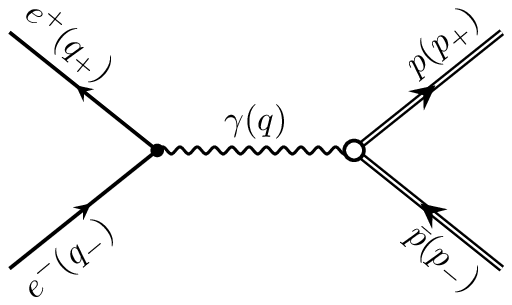}\label{fig.BornDiagram}}
	\hspace{0.05\textwidth}
    \subfigure[]{\includegraphics[width=0.4\textwidth]{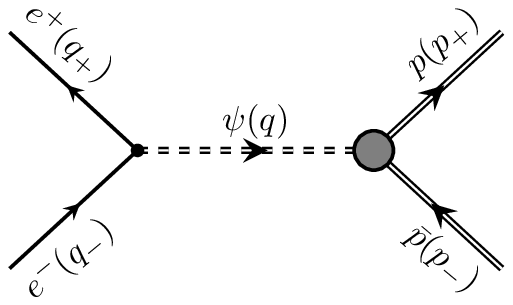}\label{fig.PsiDiagram}}
    \caption{
    	Feynman diagrams of the process $e^+e^- \to p \bar{p}$ in Born approximation (left)
    	and with the charmonium $\psi(3770)$ intermediate state (right).
    }
    \label{fig.TwoMechanisms}
\end{figure}
In Born approximation (see Fig.~\ref{fig.BornDiagram}) the electron--positron pair annihilates into
virtual photon, which then produces proton--antiproton pair. The amplitude $\M_B$ corresponding to this
process has the form:
\eq{
    \M_B = \frac{1}{s} J^{e\bar{e}\to\gamma}_\mu(q) \, J_{\gamma\to p\bar{p}}^\mu(q),
    \label{eq.BornAmplitude}
}
where $s = q^2 = \br{q_++q_-}^2 = \br{p_++p_-}^2$
is the total invariant mass squared of the lepton pair ($q$ is the momenta of intermediate photon).
The quantities $J^{e\bar{e}\to\gamma}_\mu$ and $J^{\gamma\to p\bar{p}}_\mu$
from (\ref{eq.BornAmplitude}) are lepton and proton electromagnetic currents:
\eq{
    J^{e\bar{e}\to\gamma}_\mu(q) &= -e \brs{\bar{v}(q_+) \gamma_\mu u(q_-)},
    \label{eq.CurrentEEGamma}\\
    J^{\gamma\to p\bar{p}}_\mu(q) &= e \brs{\bar{u}(p_+) \Gamma_\mu(q) v(p_-)},
    \label{eq.CurrentPPGamma}
}
where $e$ is the modulus of electron charge $e = \sqrt{4\pi\alpha}$ and $\alpha \approx 1/137$ is
the fine structure constant \cite{Zyla:2020zbs}. The proton current is described in terms of
the proton electromagnetic form factors:
\eq{
	\Gamma_\mu(q) &= F_1(q^2) \gamma_\mu - \frac{F_2(q^2)}{4M_p} \br{\gamma_\mu \dd{q} - \dd{q} \gamma_\mu}.
}
where we use the notation $\dd{a} \equiv a_\mu \gamma^\mu$.
Here $M_p$ is the mass of proton and functions $F_{1,2}(q^2)$ are the proton electromagnetic form factors
normalized as $F_1(0)=1$ and $F_2(0)=\mu_p-1$, where $\mu_p$ is the proton anomalous magnetic moment.
In paper~\cite{Ahmadov:2013ova} we used the result of the paper \cite{Ferroli:2010bi} and assumed that
energy range under our consideration, i.e. $\sqrt{s} \sim 3-4~\GeV$, is not far from
$p\bar{p}$ production threshold and thus we can neglect by $F_2$ and use the point-like proton
approximation, putting $F_1 = 1$.
However recent analysis \cite{Tomasi-Gustafsson:2020vae}
showed that the situation is not so evident and point-like proton approximation is not valid
at this energy range.
That is why we use the effective form factor G(s):
\eq{
	\brm{G(s)} = \frac{C}{s^2 \log^2\br{s/\Lambda^2}},
	\label{eq.Formfactor}
}
(as it was done in \cite{Ablikim:2014jrz} following to the results of \cite{Aubert:2005cb})
which is obtained in the assumption that
electric $G_E$ and magnetic $G_M$ form factors of the proton are equal: $\brm{G_E} = \brm{G_M}$.
That assumption leads to $F_1(s)=G(s)$ and $F_2(s)=0$. The formfactor (\ref{eq.Formfactor})
is the pQCD inspired form factor \cite{Lepage:1979za,Lepage:1980fj},
which is in a fair agreement with the cross section of the process (\ref{eq.Process})
in the energy range $\sqrt{s}$ from $2~\GeV$ to $3.07~\GeV$ measured
by BES \cite{Ablikim:2005nn}.
In equation (\ref{eq.Formfactor}) the quantity $\Lambda = 300~\MeV$ is the QCD scale and
$C$ is a free parameter fitted in \cite{Ablikim:2014jrz} to be equal to $C = (62.6 \pm 4.1)~\GeV^4$.
This fit also agrees with more recent result \cite{Bianconi:2015owa} where
this constant was fitted to be $C = 72~\GeV^4$ while $\Lambda = 520~\MeV$.
During our numerical estimations we use the BES values of $C$ and $\Lambda$.

\begin{figure}
    \centering
    \includegraphics[width=0.4\textwidth]{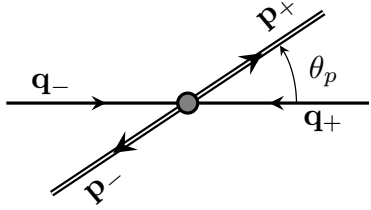}
    \caption{The definition of scattering angle $\theta_p$ from (\ref{eq.Phi2}) in
    the center-of-mass reference frame.}
    \label{fig.MomentaPosition}
\end{figure}

\begin{figure}
	\centering
	{\footnotesize	
	\input{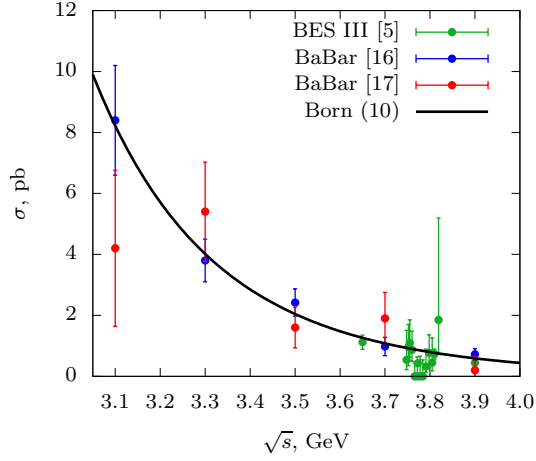}
	}
    \caption{Experimental total cross section for the process $e^+ +e^-\to \bar p+p$ around
    		  charmonium $\psi(3770)$.}
    \label{fig.WideRange}
\end{figure}

Using the amplitude $\M_B$ from (\ref{eq.BornAmplitude}) one can write down the cross section
in the standard way:
\eq{
    d\sigma_B = \frac{1}{8 s} \sum_{\text{spins}} \brm{\M_B}^2 \, d\Phi_2,
    \label{eq.BornCrossSectionGeneralForm}
}
where
the summation of the amplitude square $\brm{\M_B}^2$ runs over all possible initial and final particles spin states.
We also systematically neglect the mass of the electron $m_e$ in this paper.
The phase volume of final particles $d\Phi_2$ has the form:
\eq{
    d\Phi_2 &=
    \frac{1}{\br{2\pi}^2} \delta\br{q_++q_- - p_- - p_+} \frac{d\vv{p_+}}{2E_+}\frac{d\vv{p_-}}{2E_-}
    =
    \nn\\
    &=
    \frac{\brm{\vv{p}}}{2^4 \pi^2 \sqrt{s}} \, d\Omega_p
    =
    \frac{\beta}{2^4\pi} \, d\cos\theta_p,
    \qquad
    d\Omega_p=d\phi_p \, d\cos\theta_p,
    \label{eq.Phi2}
}
where $\phi_p$ and $\theta_p$ are the azimuthal and polar angles of the final proton in the center-of-mass
reference frame (center of mass system, c.m.s),
i.e. $\theta_p$ is the angle between 3-momenta of the initial electron $\vv{q_-}$ and the final proton
$\vv{p_+}$ (see Fig.~\ref{fig.MomentaPosition})
and $\brm{\vv{p}} \equiv \brm{\vv{p_+}} = \brm{\vv{p_-}} = \sqrt{s} \beta / 2$ is
the modulus of 3-momenta of the final proton or antiproton.
The quantity $\beta = \sqrt{1-4M_p^2/s}$ is the final proton velocity.
Using the explicit form of $\M_B$ from (\ref{eq.BornAmplitude}) and
integrating over the final particles phase space (\ref{eq.Phi2}) from (\ref{eq.BornCrossSectionGeneralForm})
one obtains:
\eq{
    \frac{d\sigma_B}{d \cos\theta_p} = \frac{\pi \alpha^2 \beta}{2s} \br{2-\beta^2\sin^2\theta_p} \brm{G\br{s}}^2.
}
The total cross section in Born approximation then reads as:
\eq{
    \sigma_B(s)=\frac{2\pi\alpha^2}{3s} \beta \br{3-\beta^2} \brm{G\br{s}}^2,
    \label{eq.TotalCrossSectionBorn}
}
which is in a good agreement with the experimental data on a wide energy range
far from the resonance $\psi(3770)$, see Fig.~\ref{fig.WideRange}.

\section{The quarkonium $\psi(3770)$ intermediate state}
\label{sec.PsiIntermediateState}

As one can see in Fig.~\ref{fig.WideRange} the total cross section including
only the electromagnetic mechanism (\ref{eq.TotalCrossSectionBorn}) fails to describe
this delicate behaviour in the vicinity of the charmonium resonance $\psi(3770)$.
Obviously in this region one should take into account the additional contribution to
the amplitude which appears from the diagram with $\psi(3770)$ in the intermediate state
(see Fig.~\ref{fig.PsiDiagram}) and is enhanced by Breit--Wigner propagator.
This leads the total amplitude $\M$ of the process (\ref{eq.Process}) to become the sum of two terms:
\eq{
    \M=\M_B+\M_\psi,
}
where $\M_B$ is the Born amplitude from (\ref{eq.BornAmplitude}) (see Fig.~\ref{fig.BornDiagram})
and $\M_\psi$ takes into account this second mechanism (see Fig.~\ref{fig.PsiDiagram}):
\eq{
    \M_\psi = \frac{1}{s-M_\psi^2+i M_\psi\,\Gamma_\psi} J^{e\bar{e}\to\psi}_\mu(q)
    \br{ g^{\mu\nu} - \frac{q^\mu q^\nu}{M_\psi^2} } J^{\psi\to p\bar{p}}_\nu (q),
    \label{eq.Mpsi}
}
where $M_\psi$ and $\Gamma_\psi$ are the mass and the total decay width of $\psi(3770)$ resonance and
$J^{e\bar{e}\to\psi}_\mu$ and $J^{\psi\to p\bar{p}}_\mu$ are the currents which describe the transition
of lepton pair into $\psi(3770)$ resonance and the transition of the $\psi(3770)$ resonance into
proton--antiproton pair correspondingly.
We notice that the second term in parenthesis in (\ref{eq.Mpsi}) does not contribute
since the currents have to be conserved:
$q^\mu J^{e\bar{e}\to\psi}_\mu = q^\mu J^{\psi\to p\bar{p}}_\mu = 0$.
Next step is to assume that $J^{e\bar{e}\to\psi}_\mu$ has the same structure as $J^{e\bar{e}\to\gamma}_\mu$ from (\ref{eq.CurrentEEGamma}), i.e.
\eq{
    J^{e\bar{e}\to\psi}_\mu(q) = g_e \, \brs{\bar{v}(q_+) \gamma_\mu u(q_-)},
}
where the constant $g_e = F_1^{\psi\to p\bar{p}}(M_\psi^2)$ is the value of the form factor of the
vertex $\psi \to p\bar{p}$ at the $\psi(3770)$ mass-shell
(here we follow the same approximation as in the Born case and
assume that $F_2^{\psi\to p\bar{p}}(M_\psi^2) = 0$).
This constant is defined via $\psi\to e^+ e^-$ decay width
$\Gamma_{\psi\to e^+e^-} = 261~\eV$ \cite{Zyla:2020zbs} which gives the following value:
\eq{
    g_e= \sqrt{\frac{12\pi\Gamma_{\psi\to e^+e^-}}{M_\psi}} = 1.6 \cdot 10^{-3}.
}
We neglect a possible imaginary part of vertex $e\bar{e}\to\psi$ since it
was shown in \cite{Kuraev:2013swa} that it is small, less then 10~\% of the real part.

The strategy for calculating the contribution $\M_\psi$ to the cross section is the
following. If we introduce the relative phase $\phi$ between Born contribution $\M_B$
and the additional contribution $\M_\psi$ then we can write the cross section as:
\eq{
	&\sigma \sim \brm{\M}^2 = \brm{\brm{\M_B} + e^{i\phi} \brm{\M_\psi}}^2
	=\nn\\
	&\qquad=
	\brm{\M_B}^2 + 2\cos\phi \brm{\M_B} \cdot \brm{\M_\psi} + \brm{\M_\psi}^2
	\sim \sigma_B + \sigma_{int} + \sigma_\psi.
	\label{eq.TotalCrossSection}
}
Knowing the Born cross section $\sigma_B$ from (\ref{eq.TotalCrossSectionBorn}) and
the interference contribution $\sigma_{int}$ with the phase $\phi$ one can calculate
the total cross section including both contributions using (\ref{eq.TotalCrossSection})
and evaluating $\sigma_\psi$ in the following manner:
\eq{
	\sigma_\psi = \br{ \frac{\sigma_{int}}{2 \cos\phi \, \sqrt{\sigma_B}} }^2.
}
Thus we need to evaluate only the interference of the resonant amplitude $\M_\psi$ with the Born amplitude $\M_B$
which has the standard form:
\eq{
    d\sigma_{int} = \frac{1}{8s} \sum_{\text{spins}} 2\,\Re\brs{\M_B^+ \M_\psi} \, d\Phi_2.
}
Our goal is to get the contribution to the total cross section thus we integrate
over the final particles phase space:
\eq{
    \sigma_{int}(s)
    &=
    \frac{1}{4 s^2} \Re\brf{
    	\frac{\sum_{s}
    	\br{J^{e\bar{e}\to\gamma}_\mu}^* J^{e\bar{e}\to\psi}_\nu}
    	{s-M_\psi^2+i M_\psi\,\Gamma_\psi}
    	\sum_{s'}
    	\int d\Phi_2
    	\br{J_{\gamma\to p\bar{p}}^\mu}^* J_{\psi\to p\bar{p}}^\nu
    	},
}
where $\sum_s$ is the summation over the spin states of initial particles
and $\sum_{s'}$ is the summation over the final particles spin states.
Next we use invariant integration trick:
\eq{
	&\sum_{s'} \int d\Phi_2 	\br{J_{\gamma\to p\bar{p}}^\mu}^* J_{\psi\to p\bar{p}}^\nu
	=
	\nn\\
	&\qquad\qquad=
	\frac{1}{3} \br{ g^{\mu\nu} - \frac{q^\mu q^\nu}{q^2} }
	\sum_{s'} \int d\Phi_2 	\br{J_{\gamma\to p\bar{p}}^\alpha}^* J^{\psi\to p\bar{p}}_\alpha.
}
Applying the conservation of the currents $J^{e\bar{e}\to\gamma}_\mu$ and $J^{e\bar{e}\to\psi}_\nu$,
recalling that
\eq{
	\sum_{s}\br{J^{e\bar{e}\to\gamma}_\mu}^* J_{e\bar{e}\to\psi}^\mu
	\approx
	-e g_e \, \Sp\brs{\dd{q_-} \gamma_\mu \dd{q_+} \gamma^\mu}
	\approx
	4 \, e \, g_e  s,
}
and using the phase volume from (\ref{eq.Phi2})
we get the following simplified expression of the interference contribution to the total cross
section:
\eq{
    \sigma_{int}(s)
    &= \frac{e g_e \beta}{48 \pi s}
    \times\nn\\
    &\times
    \Re\brf{
    	\frac{1}{s-M_\psi^2+i M_\psi\, \Gamma_\psi}
    	\int\limits_{-1}^1 d\cos\theta_p \sum_{s'}
    	\br{J_{\gamma\to p\bar{p}}^\alpha}^* J^{\psi\to p\bar{p}}_\alpha
    }.
    \label{eq.TotalCrossSectionInterference}
}
Thus we can present the interference contribution to the total cross section in the form:
\eq{
	\sigma_{int}(s)
	=
	\Re\br{ \frac{S_i(s)}{s-M_\psi^2+i M_\psi\, \Gamma_\psi} },
	\label{eq.SigmaIntViaSi}
}
where $S_i(s)$ contains all the dynamics of the transformation of charmonium into
proton-antiproton pair and has the following explicit form:
\eq{
	S_i(s)
    = \frac{e g_e \beta}{48 \pi s}
    	\int\limits_{-1}^1 d\cos\theta_p \sum_{s'}
    	\br{J_{\gamma\to p\bar{p}}^\alpha}^* J^{\psi\to p\bar{p}}_\alpha.
    \label{eq.Si}
}
The subscript index $i$ above denotes the type of mechanism of this transformation.
Since the mass of $\psi(3770)$ is higher than the threshold of $D$-meson pair
production it is natural to expect that the $D$-meson loop
will be the main mechanism in this reaction (see Fig.~\ref{fig.DD}).
However this mechanism appears to be small since the mass of $\psi(3770)$ exceeds the $D$-meson pair
production threshold slightly (one can see that $\br{M_\psi - 2M_D}/M_\psi \approx 1\%$).
Thus below, in Section~\ref{sec.ThreeGluonMechanism}, we consider also the OZI-violated three gluon mechanism (see Fig.~\ref{fig.3G})
which appears to give the dominant contribution.

\subsection{D-meson loop mechanism}
\label{sec.DMesonLoopMechanism}

\begin{figure}
	\centering
    \includegraphics[width=0.6\textwidth]{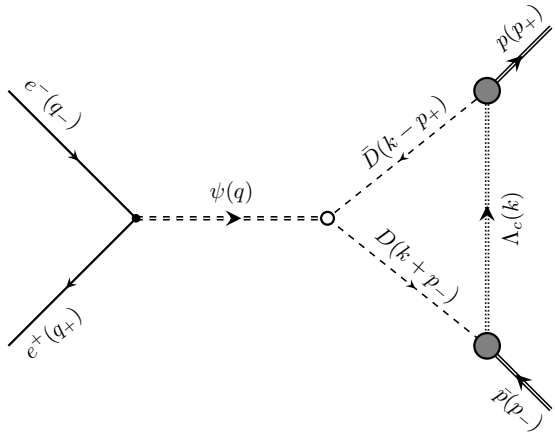}
    \caption{$D$-meson loop mechanism.}
    \label{fig.DD}
\end{figure}

The $D$-meson loop mechanism can be illustrated by the diagram in Fig.~\ref{fig.DD}.
It is known that the $\psi(3770)$ structure is somewhat complicated. For example it leads to
a specific peak shape of the cross section of the process $e^+ e^- \to D \bar{D}$
in the vicinity of $\psi(3770)$ state mass. One of the strong conjecture to interpret this fact is
to suppose the presence of a $\psi(2S)$ resonance with mass close to the $\psi(3770)$ \cite{Achasov:2012ss}.
We however do not dive into this complicated consideration since
$D$-meson loop contribution appears to be small in our case and
one can limit oneself with single two quark state with spin 1.
We introduce the following parameterizations of vertices in Fig.~\ref{fig.DD}:
\eq{
    &\includegraphics[valign=c]{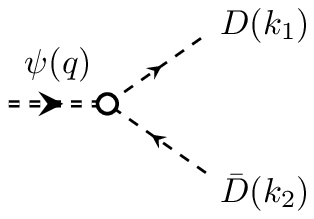}
    \quad
    \longrightarrow
    \quad
    -i G_{\psi D\bar{D}}(q^2,k_1^2,k_2^2) \br{k_1 + (-k_2)}_\mu e^\mu,
    \label{eq.VertexPsiToDD}
    \\
    \nn
    \\
    &\includegraphics[valign=c]{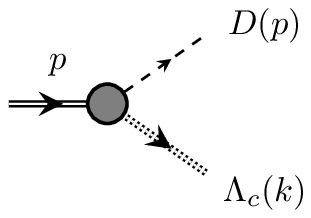}
    \quad
    \longrightarrow
    \quad
    -i G_{\Lambda D P}\br{k^2,p^2} ~ i \gamma_5,
    \label{eq.VertexPDL}
}
where $e^\mu$ is the polarization vector of charmonium $\psi(3770)$.
In the calculation we do not need to know the complete functional dependence
of functions $G_{\psi D\bar{D}}(q^2,k_1^2,k_2^2)$ and $G_{\Lambda D P}\br{k^2,p^2}$.
We discuss this dependence in this Section below.

Let us write down the $D$-meson loop contribution to the amplitude from the diagram in
Fig.~\ref{fig.DD}:
\eq{
	\M_D
	&=
    \frac{g_e}{16 \pi^2}
    \frac{\brs{ \bar{v}\br{q_+} \gamma_\mu u\br{q_-} }}{q^2-M_\psi^2+i M_\psi\,\Gamma_\psi}
    \times\nn\\
    &\qquad\times
    \int\frac{dk}{i \pi^2}
    \frac{ [\bar{u}\br{p_+} \gamma_5 (\dd{k} + M_\Lambda) \gamma_5 v\br{p_-}] \br{2 k + p_- - p_+}^\mu}
    {\br{k^2 - M_\Lambda^2}\br{(k-p_+)^2 - M_D^2} \br{(k+p_-)^2 - M_D^2}}
    \times\nn\\
    &\qquad\qquad\times G_{\psi D\bar{D}}(q^2,(k+p_-)^2,(k-p_+)^2)
    \times\nn\\
    &\qquad\qquad\times G_{\Lambda D P}\br{k^2,(k-p_+)^2} G_{\Lambda D P}\br{k^2,(k+p_-)^2},
    \label{eq.AmplitudePsiDD}
}
where $M_D$ and $M_\Lambda$ are masses of $D$-meson and $\Lambda_c$-hyperon correspondingly.
Comparing this amplitude with the general form (\ref{eq.Mpsi}) one can extract the current:
\eq{
	J^{\psi\to p\bar{p}}_\mu (q)
	&=
    \frac{1}{16 \pi^2}
    \int\frac{dk}{i \pi^2}
    \frac{ [\bar{u}\br{p_+} \gamma_5 (\dd{k} + M_\Lambda) \gamma_5 v\br{p_-}] \br{2 k + p_- - p_+}_\mu}
    {\br{k^2 - M_\Lambda^2}\br{(k-p_+)^2 - M_D^2} \br{(k+p_-)^2 - M_D^2}}
    \times\nn\\
    &\qquad\qquad\times G_{\psi D\bar{D}}(s,(k+p_-)^2,(k-p_+)^2)
    \times\nn\\
    &\qquad\qquad\times G_{\Lambda D P}\br{k^2,(k-p_+)^2} G_{\Lambda D P}\br{k^2,(k+p_-)^2},
    \label{eq.JDD}
}
and insert it into (\ref{eq.Si}). This gives the following contribution of the $D$-meson loop to the cross section
which is expressed is terms of the quantity $S_D$ from (\ref{eq.SigmaIntViaSi}):
\eq{
	S_D\br{s} &= \alpha_D\br{s} \, Z_D\br{s},
	\qquad
	\alpha_D\br{s}
	=
	\frac{\alpha~g_e}{2^4 \, 3 \pi^2} \beta\,G(s),
    \label{eq.SD}\\
	Z_D\br{s}
	&=
	\frac{1}{s}
    \int\frac{dk}{i \pi^2}
    \frac{SpD(s,k^2)}
    {\br{k^2 - M_\Lambda^2}\br{(k-p_+)^2 - M_D^2} \br{(k+p_-)^2 - M_D^2}}
    \times\nn\\
    &\qquad\qquad\times G_{\psi D\bar{D}}(s,(k+p_-)^2,(k-p_+)^2)
    \times\nn\\
    &\qquad\qquad\times G_{\Lambda D P}\br{k^2,(k-p_+)^2} G_{\Lambda D P}\br{k^2,(k+p_-)^2},
    \label{eq.ZD}
}
where $SpD(s,k^2)$ is the trace of $\gamma$-matrices over the baryon line:
\eq{
	SpD(s,k^2)
	&=
    \Sp\brs{(\dd{p_+}+M_p) \gamma_5 (\dd{k}+M_\Lambda) \gamma_5 (\dd{p_-}-M_p) (\dd{k} - M_p)}
    =\nn\\
    &=
    2\brs{
    	\br{k^2}^2
    	+
    	k^2\br{
    		s - 2\br{M_D^2 + M_p M_\Lambda}
    	}
    	-
    	s M_p M_\Lambda
    	+
    	c_D
    },
    \label{eq.SpDExplicit}
    \\
    c_D
    &=
    M_D^4 + 2 M_p M_\Lambda M_D^2 + 2 M_\Lambda M_p^3 - M_p^4.
    \label{eq.cD}
}
Now we need to evaluate the quantity $Z_D\br{s}$ from (\ref{eq.ZD}).
In order to do this we use Cutkosky rule \cite{Cutkosky:1961} for $D$-meson propagators:
\eq{
	&\frac{1}{(k+p_-)^2 - M_D^2}
	\quad\longrightarrow\quad
	- 2 \pi i ~ \delta\br{(k+p_-)^2 - M_D^2} ~\theta\br{(k+p_-)_0},
	\nn\\
	&\frac{1}{(k-p_+)^2 - M_D^2}
	\quad\longrightarrow\quad
	- 2 \pi i ~ \delta\br{(k-p_+)^2 - M_D^2} ~\theta\br{-(k-p_+)_0},
	\nn
}
and obtain the imaginary part of this quantity:
\eq{
	2i \, \Im \, Z_D\br{s}
	&=
	\frac{\br{- 2\pi i}^2}{s}
    \int\frac{dk}{i \pi^2}
    \frac{SpD} {k^2 - M_\Lambda^2}
    G_{\psi D\bar{D}}(s,(k+p_-)^2,(k-p_+)^2)
    \times\nn\\
    &\qquad\times
    G_{\Lambda D P}\br{k^2,(k-p_+)^2}~G_{\Lambda D P}\br{k^2,(k+p_-)^2}
    \times\nn\\
    &\qquad\times
    \delta\br{(k+p_-)^2 - M_D^2}~\delta\br{(k-p_+)^2 - M_D^2}
    \times\nn\\
    &\qquad\times
    \theta\br{(k+p_-)_0}~\theta\br{-(k-p_+)_0}.
    \label{eq.ImZD0}
}
We notice that one can utilize these two $\delta$-functions in (\ref{eq.ImZD0})
to significantly simplify the evaluation of $\Im \, Z_D$. For example,
one can see that usage of $\delta$-functions leads to
the replacements:
\eq{
	\delta((k-p_+)^2 - M_D^2)
	\qquad\rightarrow\qquad
	2\br{kp_+} = k^2 + M_p^2 - M_D^2,
	\\
	\delta((k+p_-)^2 - M_D^2)
	\qquad\rightarrow\qquad
	2\br{kp_-} = M_D^2 - M_p^2 - k^2,
}
which we already used in (\ref{eq.SpDExplicit}).
Performing the loop integrations one gets:
\eq{
	\Im \, Z_D\br{s}
    &=
	-\frac{2\pi}{s^{3/2}}
	G_{\psi D\bar{D}}(s,M_D^2,M_D^2)
	\int\limits_{C_k^{(1)}}^1 \frac{dC_k}{\sqrt{D_1}}
    \sum_{i=1,2}
    \frac{k_{(i)}^2} {k_{(i)}^2 + M_\Lambda^2}~
	\times\nn\\
	&\quad\times
    SpD(s,-k_{(i)}^2)~G_{\Lambda D P}^2\br{-k_{(i)}^2,M_D^2},
    \qquad
	s > 4M_D^2,
    \label{eq.ImZD}
}
with the integration over cosine of polar angle $C_k = \cos\theta_k$ is evaluated
numerically below.
\ref{app.IntegrateUsingDeltas} details the derivation of eq.~(\ref{eq.ImZD}) and the
definitions of the quantities $C_k^{(1)}$, $k_{(i)}$ and $D_1$.

Now we consider the explicit expression of form factors which we need to evaluate
$\Im \, Z_D$ from (\ref{eq.ImZD}). First we see that for $\psi \to D\bar{D}$ vertex we
need only dependence over charmonium virtuality $q^2 = s$, since $D$-meson legs are on-mass-shell.
We can start with the normalization of function $G_{\psi D\bar{D}}(s,M_D^2,M_D^2)$
to the decay of charmonium $\psi(3770)$ into $D\bar{D}$ final state.
Calculating this decay width one gets:
\eq{
	g_{\psi D\bar{D}}
	\equiv
	G_{\psi D\bar{D}}(M_\psi^2,M_D^2,M_D^2)
	=
	4\,\sqrt{\frac{3\pi \, \Gamma_{\psi \to D\bar{D}}}{M_\psi \, \beta_D^3}}
	=
	18.4,
	\label{eq.gPsiDD}
}
where $\beta_D = \sqrt{1 - 4M_D^2/M_\psi^2}$ is the $D$-meson velocity in this decay.
The numerical value of $g_{\psi D\bar{D}}$ in (\ref{eq.gPsiDD}) is obtained by
using the experimental value of the width of charmonium decay into charged ($D^+ D^-$) and
neutral ($D_0 \bar{D}_0$) mesons
$\Gamma_{\psi \to D\bar{D}} = 25~\MeV$ \cite{Zyla:2020zbs}.
This let us to take into account all these types of $D$-mesons in the loop in Fig.~\ref{fig.DD}.

The dynamical dependence of function $G_{\psi D\bar{D}}(s,M_D^2,M_D^2)$
we propose following to \cite{Lepage:1979zb} in the form:
\eq{
	G_{\psi D\bar{D}}\br{s,M_D^2,M_D^2} = \frac{C_{\psi D\bar{D}}}{s \, \log\br{s/\Lambda_D^2}},
	\qquad
	s > 0,
	\label{eq.PsiDDFormfactor0}
}
where constant $C_{\psi D\bar{D}}$ can be found from the normalization (\ref{eq.gPsiDD})
and we get $C_{\psi D\bar{D}} = g_{\psi D\bar{D}} \, M_\psi^2 \log\br{M_\psi^2/\Lambda_D^2}$.
Finally we use the following explicit form of function:
\eq{
	G_{\psi D\bar{D}}\br{s,M_D^2,M_D^2}
	=
	g_{\psi D\bar{D}} \, \frac{M_\psi^2}{s} \, \frac{\log\br{M_\psi^2/\Lambda_D^2}}{\log\br{s/\Lambda_D^2}},
	\label{eq.PsiDDFormfactor}
}
where scale $\Lambda_D$ we fix on the characteristic value of the reaction $\Lambda_D = 2 M_D$.

Next we consider function $G_{\Lambda D P}\br{k^2,p^2}$ from (\ref{eq.ImZD}). And again
the only dependence left after the application of Cutkosky rule is
the off-mass-shellness of $\Lambda_c$-hyperon
in the scattering regime, since here $k^2 = -k_{(i)}^2 < 0$.
The dependence over the virtuality of the $D$-meson has disappeared,
$D$-mesons are on mass shell. But still we need to take
into account the remnants of this dependence.
Following to \cite{Ahmadov:2013ova} we use this form for
$\Lambda D P$-vertex with the off-mass-shell $D$-meson which was established
in \cite{Reinders:1984sr,Navarra:1998vi}:
\eq{
	g_D(p^2) = \frac{2 M_D^2 f_D}{m_u + m_c}
	\frac{g_{DN\Lambda}}{p^2 - M_D^2},
}
where $f_D \approx 180~\MeV$. For quark masses the following values are used:
$m_u \approx 280~\MeV$ and  $m_c = 1.27~\GeV$ \cite{Zyla:2020zbs}.
The constant $g_{DN\Lambda} \approx 6.74$ was estimated in \cite{Navarra:1998vi}
in scattering regime, i.e. for $p^2 < 0$.
Thus in our calculation in eq.~(\ref{eq.ImZD}) we use the following expression:
\eq{
	G_{\Lambda D P}(-k_{(i)}^2, M_D^2) = \frac{f_D \, g_{DN\Lambda}}{m_u + m_c},
}
which do not take the effects of the $\Lambda_c$-hyperon off-mass-shellness.
We expect that these effects are not very important.

Now we're able to calculate the imaginary part of $Z_D$ following to eq.~(\ref{eq.ImZD}).
The real part of this quantity is restored by using the dispersion relation
(see \ref{app.DispersionRelation}).

\subsection{Three gluon mechanism}
\label{sec.ThreeGluonMechanism}

\begin{figure}
	\centering
    \includegraphics[width=0.6\textwidth]{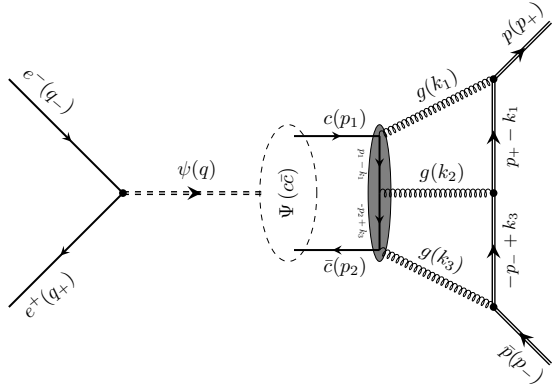}
    \caption{Three gluon mechanism.}
    \label{fig.3G}
\end{figure}

The three gluon mechanism was considered in details in \cite{Ahmadov:2013ova} and thus we
just briefly recall a few steps of this calculation in order to correct some misprints and minor
mistakes in formulas of \cite{Ahmadov:2013ova} which do not affect the conclusions of the paper.

The three gluon mechanism presented in Fig.~\ref{fig.3G} gives the following contribution
to the quantity $S_{3g}$ from (\ref{eq.SigmaIntViaSi})
(which coincide with equations (16) and (17) from \cite{Ahmadov:2013ova}):
\eq{
    S_{3g}\br{s}
    &=
    \alpha_{3g} \br{s} \, Z_{3g}\br{s},
    \qquad\quad
    \alpha_{3g}\br{s} = \frac{\alpha\, \alpha_s^3}{2^3 \, 3} g_e \, g_{col} \, \phi \, \beta \, G\br{s} \, G_\psi(s),
    \label{eq.alpha3g}
    \\
    Z_{3g}\br{s}
    &=
    \frac{4}{\pi^5 s}
    \int \frac{dk_1}{k_1^2}\frac{dk_2}{k_2^2}\frac{dk_3}{k_3^2}
    \frac{Sp3g ~ \delta\br{q-k_1-k_2-k_3}}{ (\br{p_+ - k_1}^2 - M_p^2) (\br{p_- - k_3}^2 - M_p^2) },
    \label{eq.Z3g}
}
where the quantity $Sp3g$ is the product of traces over proton and $c$-quark lines:
\eq{
	&Sp3g
	=
    \Sp \brs{ \hat{Q}_{\alpha\beta\gamma} (\dd{p_1} + m_c) \gamma^\mu (\dd{p_2} - m_c) }
    \times\nn\\
    &~~\times
   	\Sp\brs{ (\dd{p_+}+M_p) \gamma^\alpha (\dd{p_+} - \dd{k_1} + M_p) \gamma^\beta (-\dd{p_-} + \dd{k_3} + M_p) \gamma^\gamma (\dd{p_-}-M_p) \gamma_\mu },
   	\nn
}
with
\eq{
    \hat{Q}_{\alpha\beta\gamma}
    &=
    \frac{\gamma_\gamma (-\dd{p_2} + \dd{k_3} + m_c) \gamma_\beta (\dd{p_1} - \dd{k_1} + m_c) \gamma_\alpha}
    { ((p_2-k_3)^2 - m_c^2) ((p_1-k_1)^2 - m_c^2) }
    + \brs{\text{gluon permutations}},
    \label{eq.DefinitionHatQ}
}
where the permutations over gluon vertices are performed in the gray block in Fig.~\ref{fig.3G}.

First we notice that there is a misprint in second proton propagator denominator in
eq.~(18) in \cite{Ahmadov:2013ova} which is corrected here in (\ref{eq.Z3g}).
This misprint leads to the following mistakes in angular integrals in eq.~(21) and in
the Appendix~B of \cite{Ahmadov:2013ova}. The angular integration
must be performed over $dc_1 dc_3$ instead of $dc_1 dc_2$.

Next we also note that the quantity $\phi$ in $\alpha_{3g}(s)$ in (\ref{eq.alpha3g}) is similar to the quantity
$R$ from eq.~(17) of \cite{Ahmadov:2013ova}. The difference comes from the calculations of
$\psi \to 3 g$ decay which we use for the normalization of the decay constant $\phi$.
In paper \cite{Ahmadov:2013ova} (see Appendix~A there) the factor $\sqrt{2/3}$ was missed
in the amplitude of $\psi \to 3 g$ decay (see eq.~(A.7) there).
This factor comes from missed $2$ in the right-hand-side of the sum over spins in
the expression below eq.~(A.6) in \cite{Ahmadov:2013ova} and from the complete missing of
symmetrization factor $1/\sqrt{3}$ in the color wave function of charmonium
$\br{\bar{q}_1 q_1 + \bar{q}_2 q_2 + \bar{q}_3 q_3}/\sqrt{3}$ which is necessary for
correct normalization of this wave function (see, for example, eq.~(5) in \cite{Chiang:1993qi}).
Thus, after correction of this missing factor one gets:
\eq{
	\phi = \frac{\brm{\psi\br{\vv{r}=\vv{0}}}}{M_\psi^{3/2}} = \frac{R}{\sqrt{2/3}} =
    \frac{\alpha_s^{3/2}}{3\sqrt{3\pi}}.
}

\begin{figure}
	\centering
    \subfigure[]{\includegraphics[width=0.25\textwidth]{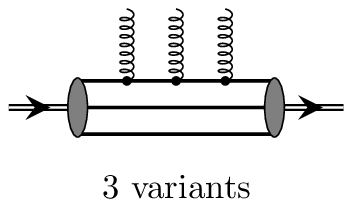}}
	\hspace{0.05\textwidth}
    \subfigure[]{\includegraphics[width=0.25\textwidth]{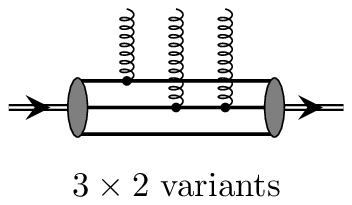}}
	\hspace{0.05\textwidth}
    \subfigure[]{\includegraphics[width=0.25\textwidth]{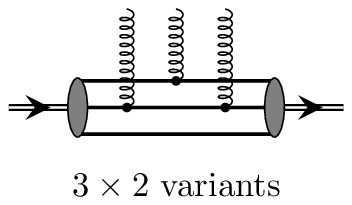}}
    \\
    \subfigure[]{\includegraphics[width=0.25\textwidth]{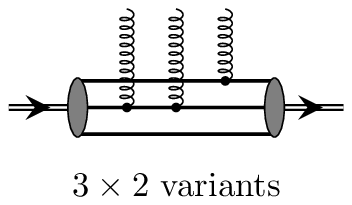}}
	\hspace{0.05\textwidth}
    \subfigure[]{\includegraphics[width=0.25\textwidth]{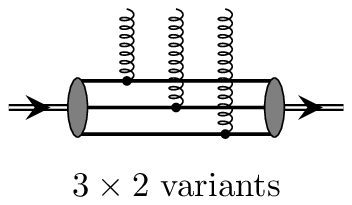}}
    \caption{
    	Possible connections of three gluons with quarks in the proton.
    }
    \label{fig.GluonsToProton}
\end{figure}

A correction should be also applied to the color factor $g_{col}$ in (\ref{eq.alpha3g}).
This factor was evaluated in eq.~(11) of
Ref.~\cite{Ahmadov:2013ova}, taking into account the summation
over gluon and quark colors. But the probability of gluon connection with the proton is evaluated by simple
multiplication of a factor of $3$ which comes from 3 quarks in the proton.
Taking into account all the possibilities (see Fig.~\ref{fig.GluonsToProton}) we get a factor of 27:
\eq{
	g_{col}
	=
	\frac{1}{4} \bra{p}  d^{ijk} ~ T^i T^j T^k \ket{p}
	=
	\frac{27}{4} \bra{q} \frac{10}{9} I \ket{q}
	=
	\frac{15}{2}.
}

The most important correction concerns the final proton--anti\-pro\-ton state.
The three gluons obtained from $\psi(3770)$ decay
produce three quark--antiquark pairs and it is implicitly assumed that they form the proton--antiproton final state.
The details of this process is not touched in Ref.~\cite{Ahmadov:2013ova} and it is mostly not important
for the relative phase.
However here we want to reproduce the absolute value of the cross section and thus we need to implement this
mechanism somehow. In general it is the mechanism of transition of three gluons (with total angular momentum equal to $1$)
into final proton--antiproton pair. We suggest that this mechanism has much in common with proton--antiproton pair
production from the photon, i.e. the electromagnetic vertex $\gamma^* \to p\bar{p}$ in time-like region.
Thus we insert into (\ref{eq.alpha3g}) the extra form factor, similar to (\ref{eq.Formfactor}),
but with some other value of the parameter $C_\psi$:
\eq{
	\brm{G_\psi(s)} = \frac{C_\psi}{s^2 \log^2\br{s/\Lambda^2}}.
	\label{eq.FormfactorPsi}
}
We find the value of this parameter $C_\psi$ in Section~\ref{sec.Numerical}.
The parameter $\Lambda$ is still the QCD scale parameter since it takes into account
the running of $\alpha_s$ coupling at microscopic scale.

The strategy of the calculation of the quantity $Z_{3g}(s)$ is the same as for the $D$-meson loop contribution
in the previous section: first we calculate its imaginary part $\Im \, Z_{3g}(s)$ by the use of Cutkosky rule
for gluon propagators:
\eq{
	\frac{1}{k_1^2 + i 0} \frac{1}{k_2^2 + i 0} \frac{1}{k_3^2 + i 0}
	\longrightarrow
	(-2\pi i)^3 \, \delta(k_1^2) \, \delta(k_2^2) \, \delta(k_3^2) \, \theta(k_1^0) \, \theta(k_2^0) \, \theta(k_3^0),
}
then we restore the real part $\Re \, Z_{3g}(s)$ by using dispersion relation technic. This strategy
is the same as in Ref.~\cite{Ahmadov:2013ova}. See all the details of these calculation there.

\section{Numerical results}
\label{sec.Numerical}

\begin{figure}
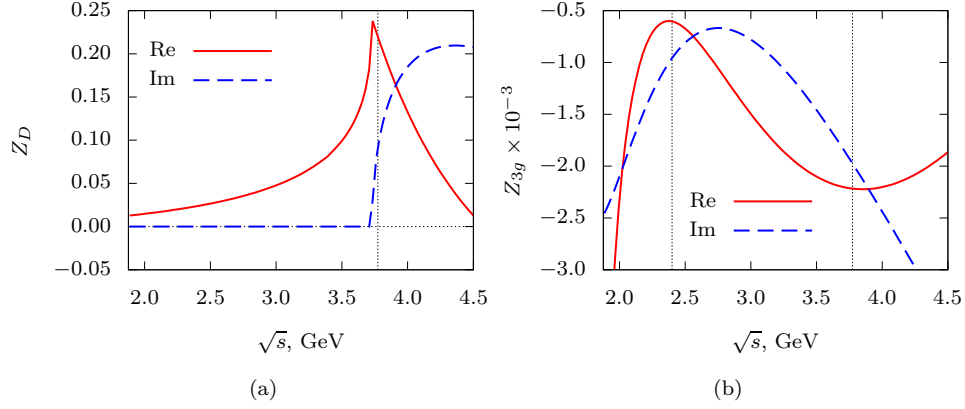

	\centering
    \subfigure[]{{\footnotesize \input{Fig7a.tex}} \label{fig.ZD}}
    ~
    \subfigure[]{{\footnotesize \input{Fig7b.tex}} \label{fig.Z3g}}
    \caption{
    	The quantities $Z_D\br{s}$ from (\ref{eq.ZD})
    	and
    	$Z_{3g}\br{s}$ from (\ref{eq.Z3g}) as a function of
    	the total invariant energy $\sqrt{s}$ starting from
    	the threshold $\sqrt{s} = 2M_p$.
    	The vertical dashed line shows the position of $\psi(3770)$.
    }
    \label{fig.ZiVsS}
\end{figure}

\begin{figure}
	\centering
    \subfigure[]{{\footnotesize \input{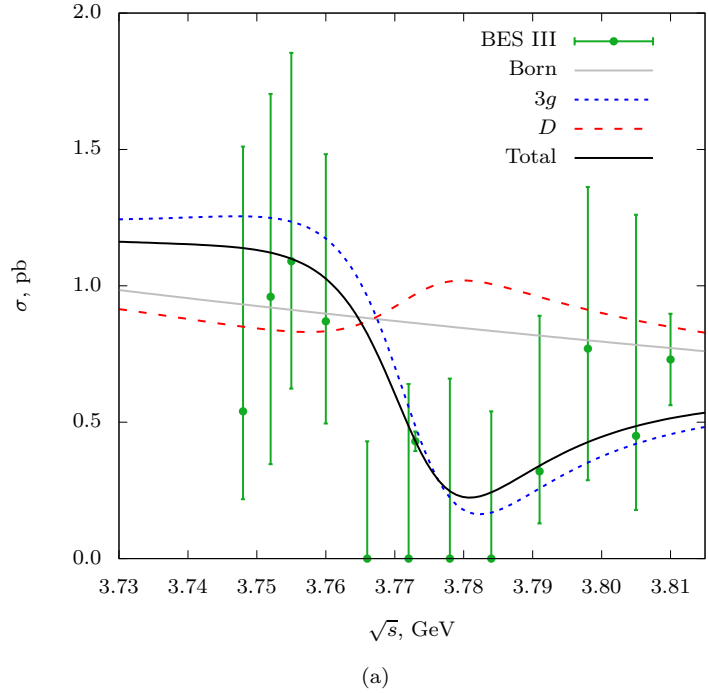}} \label{fig.Narrow}}
    \caption{
    	The total cross section (\ref{eq.TotalCrossSection}) and its different contributions
    	in the vicinity of $\psi(3770)$ resonance. The data are from the BESIII Collaboration \cite{Ablikim:2014jrz}.
    }
    \label{fig.CSNarrow}
\end{figure}

\begin{figure}
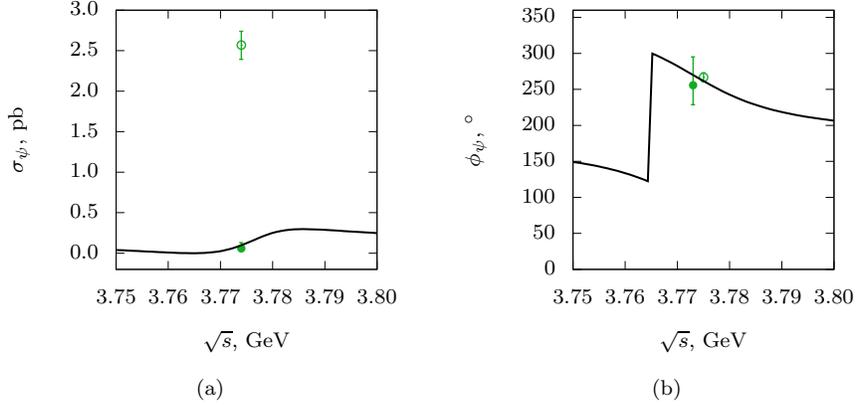

	\centering
    \subfigure[]{{\footnotesize \input{Fig9a.tex}} \label{fig.BESIIICS}}
    \subfigure[]{{\footnotesize \input{Fig9b.tex}} \label{fig.BESIIIPhase}}
    \caption{
    	The comparison with the fit of BESIII \cite{Ablikim:2014jrz}.
        The solid points correspond to the Solution~1 in Table~2 in \cite{Ablikim:2014jrz},
        while the hollow points correspond to the Solution~2.
    }
    \label{fig.CompareWithBESIIIFit}
\end{figure}

First we present the main ingredient of our calculation, the quantities $Z_D\br{s}$ from (\ref{eq.ZD})
and $Z_{3g}\br{s}$ from (\ref{eq.Z3g}) as a function of total energy $\sqrt{s}$ in the range starting from the
threshold of the reaction $s = 4M_p^2$ up to $4.5~\GeV$, see Fig.~\ref{fig.ZD} and Fig.~\ref{fig.Z3g}.
The position of the $\psi(3770)$ resonance is marked by a vertical dashed line.
One can see that the quantity $Z_{3g}$ is much bigger than $Z_D$ and both have large real and imaginary parts
in that region. The imaginary part of $Z_D$ below $D\bar{D}$ threshold
(i.e. at $s < 4M_D^2$) is zero and grows slightly up from this point with energy giving
small contribution to the observables.


Next we plot the cross section itself and compare it with
the data from the BESIII collaboration
presented in \cite{Ablikim:2014jrz}.
In Fig.~\ref{fig.CSNarrow} we present experimental data of BESIII scan of total cross section
of the reaction $e^+ e^- \to p\bar{p}$ around the mass of $\psi(3770)$ resonance and
with the specific precise measurement of the point at exactly the mass of $\psi(3770)$.
We plot the Born cross section according to (\ref{eq.TotalCrossSectionBorn}) with the
proton  electromagnetic form factor $G(s)$ from (\ref{eq.Formfactor})
as a gray line in Fig.~\ref{fig.CSNarrow}. We see that it
overestimates the precise point at $\psi(3770)$ resonance mass.
Adding the $D$-meson loop contribution (dashed line) does not reproduce the desired $\psi(3770)$ resonance point.
But if one replaces the $D$-meson loop contribution with the three-gluon mechanism
(dotted line) then the tendency of the curve becomes rather similar to the data.
And if we take into account all three contributions (solid line) then we see that our total curve
reproduce the precise $\psi(3770)$ resonance point and
follows the side shoulders, i.e. the experimental points that seem to be higher below the $\psi(3770)$
and lower above the $\psi(3770)$ resonance.
In fact this coincidence of our total curve with the precise data point is the result of
fitting of the constant $C_\psi$ from (\ref{eq.FormfactorPsi}) to the best measured experimental
point at $\psi(3770)$ mass.
Since the dependence of the total cross section of the constant $C_\psi$
is quadratic we have two solutions for it. Using the data point errorbars we
can estimate that the range of the values of this constant should be
$33.7~\GeV^4 < C_\psi < 59.1~\GeV^4$ with the best fit to the central point at two values:
\eq{
	\text{Fit 1:}	\qquad C_\psi &= 42.1~\GeV^4,	\label{eq.Fit1}
	\\
	\text{Fit 2:}	\qquad C_\psi &= 50.6~\GeV^4.	\label{eq.Fit2}
}
We find this values to be rather close to the one from the electromagnetic
vertex case, see the value of parameter $C$ below eq.~(\ref{eq.Formfactor}).
In all our numerical estimations we use the value of Fit~1 from
(\ref{eq.Fit1}). The usage of Fit~2 gives the slight change in the plot shoulders
but keeps the central point untouched.

Now we can compare our calculation with the fit of the expression
for the total cross section (1) in \cite{Ablikim:2014jrz}:
\eq{
    \sigma(s)
    =
    \brm{
        \sqrt{\sigma_B(s)}
        +
        \sqrt{\sigma_\psi}~
        \frac{M_\psi \Gamma_\psi}{s - M_\psi^2 + i M_\psi \Gamma_\psi}~
        e^{i \phi_\psi}
    }^2,
    \label{eq.FittingForm}
}
where the quantities $\sigma_\psi$ and $\phi_\psi$ are considered as parameters to be
fitted on the data.
The fitting procedure gave two solutions (see Table~2 in \cite{Ablikim:2014jrz}):
\eq{
    &\text{Solution 1}:
    \quad
    &\sigma_\psi &= \br{0.059^{+0.070}_{-0.020} \pm 0.012}~\pb,
    \quad
    &\phi_\psi &= \br{255.8^{+39.0}_{-26.6} \pm 4.8}^{\circ},
    \nn\\
    &\text{Solution 2}:
    \quad
    &\sigma_\psi &= \br{2.57^{+0.12}_{-0.13} \pm 0.12}~\pb,
    \quad
    &\phi_\psi &= \br{266.9^{+6.1}_{-6.3} \pm 0.9}^{\circ}.
    \nn
}
We plot these two solutions in Fig.~\ref{fig.CompareWithBESIIIFit} at the $\psi(3770)$ resonance mass
(in Fig.~\ref{fig.BESIIIPhase} we shifted the points slightly aside from $\psi(3770)$ resonance mass
to make the points seen separately).
Assuming the form (\ref{eq.FittingForm}) we extract the quantities $\sigma_\psi$ and $\phi_\psi$
from our result for total cross section (\ref{eq.TotalCrossSection}) and plot them in
Fig.~\ref{fig.CompareWithBESIIIFit}.
One can see that both phases are consistent with our curve
(see Fig.~\ref{fig.BESIIIPhase}) while only one value for $\sigma_\psi$ agrees with our result
(see Fig.~\ref{fig.BESIIICS}).
Namely the solution 1 from Table~2 in \cite{Ablikim:2014jrz} agrees with our numbers:
\eq{
    \sigma_\psi = 0.075~\pb,
    \qquad
    \phi_\psi = 270^{\circ}.
}

Finally we note that our calculation of three gluon mechanism contains the QCD coupling
constant $\alpha_s$ at charmonium scale (i.e. at $s \sim M_c^2$)
in rather high degree (see eq.~(\ref{eq.alpha3g})) and thus it is very sensitive
to its value.
We use the value $\alpha_s(M_c) = 0.28$ which is expected
by the QCD evolution of $\alpha_s$ from the $b$-quark scale to the
$c$-quark scale. We should note that this value differs from
the one for $J/\psi$ charmonium for which one must use much smaller value
of parameter $\alpha_s(M_c) = 0.19$
\cite{Chiang:1993qi}.

\section{Conclusion}
\label{sec.Conclusion}

We considered the process of electron--positron annihilation into
proton--antiproton pair in the vicinity of charmonium $\psi(3770)$ resonance.
Besides the Born mechanism, which is the pure QED, there are two contributions
related with intermediate charmonium $\psi(3770)$ state.
One of them is the $D$-meson loop and the other is three gluon mechanism.

We showed that $D$-meson loop mechanism, being the most probable candidate
to describe this process, fails to reproduce the value and the shape of the cross section.

It has been shown that the main contribution comes from three gluon mechanism which
can reproduce the position of the precise experimental point at the $\psi(3770)$ resonance mass
and also grasps the shape of the curve shoulders around it.

Having all the calculation in the hands we are able to select one of the two fit
solutions obtained in \cite{Ablikim:2014jrz} and to provide a solid basis for the
consideration of similar processes with binary final states and with charmonium in the
intermediate state.
This will be the subject of our future works.

\section*{Acknowledgements}

The author wish to express his gratitude
to Dr.~V.A.~Zykunov for intensive and valuable discussions,
to Dr.~A.E.~Dorokhov for reading the manuscript and for important criticism and
to Dr.~E.~Tomasi-Gustafsson for careful reading and valuable comments.
The author is also grateful to professor~V.S.~Fadin for
important comment on the paper \cite{Ahmadov:2013ova} during the
discussion in 2013, which lead to the significant correction of our approach.

\appendix

\section{Integration over $dk$ in (\ref{eq.ImZD0}) using $\delta$-functions}
\label{app.IntegrateUsingDeltas}

\begin{figure}
    \centering
    \includegraphics[width=0.4\textwidth]{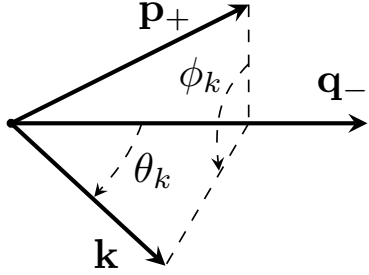}
    \caption{The definition of the polar angle $\theta_k$ and the azimuthal angle $\phi_k$
    for loop momentum $\vv{k}$ in (\ref{eq.A.init}).}
    \label{fig.LoopMomentaOrientation}
\end{figure}

In this section we show how one can obtain (\ref{eq.ImZD}) from (\ref{eq.ImZD0})
utilizing the benefits of $\delta$-functions in the integrand.
In order to do this we consider the following expression:
\eq{
    I =&\int\limits_{-1}^1 d\cos\theta_p
    \int\limits_{-\infty}^\infty dk_0
    \int\limits_0^\infty k^2 dk
    \int\limits_{-1}^1 dC_k
    \int\limits_0^{2\pi} d\phi_k
    ~F(k_0, k)
    \times\nn\\
    &\qquad\qquad\times
	\delta\br{(k+p_-)^2 - M_D^2}~\delta\br{(k-p_+)^2 - M_D^2}
    \times\nn\\
    &\qquad\qquad\times
    \theta\br{(k+p_-)_0}~\theta\br{-(k-p_+)_0},
	\label{eq.A.init}
}
where $C_k \equiv \cos\theta_k$, $\theta_k$ and $\phi_k$ are the polar and
azimuthal angles of loop momentum $\vv{k}$ which are measured from the direction of
final proton momentum $\vv{p}_+$ (see Fig.~\ref{fig.LoopMomentaOrientation}).
In this section we use the notation $k \equiv \brm{\vv{k}}$.
The function $F(k_0, k)$ contains all non-trivial part of the integrand from (\ref{eq.ImZD0}):
\eq{
	F(k_0, k) =
	\frac{SpD(s,k_0^2-k^2)}
	{k_0^2-k^2-M_\Lambda^2} G_{\Lambda D P}^2(k_0^2-k^2, M_D^2).
}
First we transform the second $\delta$-function in (\ref{eq.A.init}) to the form:
\eq{
	\delta((k-p_+)^2 - M_D^2) ~\theta\br{(k+p_-)_0}
	=
	\frac{1}{\sqrt{D}} \delta( k_0 - k_0^{(1)} ),
	\label{eq.A.delta1}
}
where the quantity $k_0^{(1)}$ is one of the poles of the argument of $\delta$-function
with respect to the variable $k_0$:
\eq{
	k_0^{(1)} = \frac{1}{2}\br{ \sqrt{s} - \sqrt{D} },
	\qquad
	D = s + 4\br{k^2 - \sqrt{s} \, \beta \, k \, C_k - M_p^2 + M_D^2}.
}
The remaining $\delta$-function can be transformed into the following form:
\eq{
	\left.\delta((k+p_-)^2 - M_D^2) ~\theta\br{(k+p_-)_0}\right|_{k_0 \to k_0^{(1)}}
	=
	\frac{1}{2\sqrt{D_1}} \sum_{i=1,2} \delta( k - k_{(i)} ),
	\label{eq.A.delta2}
}
where $k_{(i)}$ are the poles of the argument of $\delta$-function
with respect to the variable $k$:
\eq{
	k_{(1,2)} = \frac{1}{2} \br{
		\sqrt{s} \, \beta \, C_k \pm \sqrt{ D_1 }
	},
	\qquad
	D_1 = s \, \beta^2 \, C_k^2 - 4\br{M_D^2 - M_p^2}.
}
We note that if $k=k_{(1,2)}$ then $D=s$ and thus $k_0 = k_0^{(1)} = 0$.
The requirement that $D_1 > 0$ leads to the restriction
that $C_k > C_k^{(1)}$ or $C_k < C_k^{(2)}$, where:
\eq{
	C_k^{(1,2)} = \pm \frac{2}{\beta} \sqrt{\frac{M_D^2 - M_p^2}{s}},
}
while $k_{(1,2)} > 0$ only for $C_k > C_k^{(1)}$. Gathering all this together
we can integrate over $dk_0$ using $\delta$-function (\ref{eq.A.delta1})
and over $dk$ using $\delta$-function (\ref{eq.A.delta2}) in (\ref{eq.A.init})
and obtain:
\eq{
	I
    =
	\frac{1}{2\sqrt{s}}
    \int\limits_{-1}^1 d\cos\theta_p
    \int\limits_0^{2\pi} d\phi_k
    \int\limits_{C_k^{(1)}}^1 \frac{dC_k}{\sqrt{D_1}}
    \,
	\sum_{i=1,2} k_{(i)}^2 ~F(0, k_{(i)}),
}
Finally, we notice that integration over $d\cos\theta_p$
and $d\phi_k$ is trivial and gives $4\pi$:
\eq{
	I =
	\frac{2\pi}{\sqrt{s}}
    \int\limits_{C_k^{(1)}}^1 \frac{dC_k}{\sqrt{D_1}}
    \,
	\sum_{i=1,2} k_{(i)}^2 ~F(0, k_{(i)}),
	\qquad
	s > 4M_D^2,
	\label{eq.A.last}
}
where the threshold condition ($s > 4M_D^2$) comes from the fact that the integral
is not zero only if $C_k^{(1)} < 1$.
Comparing (\ref{eq.A.init}) and (\ref{eq.A.last}) we prove that
(\ref{eq.ImZD}) follows from (\ref{eq.ImZD0}).

\section{Dispersion relations}
\label{app.DispersionRelation}

In this section we derive dispersion relations to restore the real parts of
the quantities $Z_D$ from (\ref{eq.ZD}) and $Z_{3g}$ from (\ref{eq.Z3g}).
We apply dispersion relations over the variable $s$ with subtraction
at the point $s = 0$:
\eq{
	\Re \, Z_i(s) = \Re \, Z_i(0) + \frac{s}{\pi} {\cal P}
	\int\limits_{s_{\text{min}}}^\infty \frac{ds_1}{s_1 \br{s_1-s}} ~\Im \, Z_i(s_1),
	\label{eq.B.init}
}
where $s_{\text{min}}$ is the minimal threshold at which $\Im \, Z_i(s_1)$ becomes non-zero.
The substraction constant $\Re \, Z_i(0)$ vanishes (i.e. $\Re \, Z_D(0)=0$)
since there are no open charm in the proton and thus in the Compton limit the vertex
$\psi \to p\bar{p}$ must vanish.

Next we substitute the variable $\beta$:
\eq{
	\beta &= \sqrt{1 - \frac{4M_p^2}{s}},
	\quad
	\longrightarrow
	\quad
	s = \frac{4M_p^2}{1-\beta^2},
	\nn\\
	\beta_1 &= \sqrt{1 - \frac{4M_p^2}{s_1}},
	\quad
	\longrightarrow
	\quad
	s_1 = \frac{4M_p^2}{1-\beta_1^2},
	\quad
	\longrightarrow
	\quad
	ds_1 = \frac{8 M_p^2 \beta_1}{\br{1-\beta_1^2}^2} d\beta_1.
	\nn
}
Substituting these replacements into (\ref{eq.B.init}) one gets:
\eq{
	\Re \, Z_i(\beta)
	&=
	\frac{{\cal P}}{\pi} \int\limits_{\beta_{\text{min}}}^1
	\frac{2\beta_1 d\beta_1}{\beta_1^2 - \beta^2}~ \Im \, Z_i(\beta_1),
}
where $\beta_{\text{min}} = \sqrt{1 - 4 M_p^2 / s_{\text{min}}}$.
In order to improve the numerical stability of this integral we transform it:
we add and subtract the regular part of the numerator in the point $\beta$:
\eq{
	\Re \, Z_i(\beta)
	&=
	\frac{{\cal P}}{\pi} \int\limits_{\beta_{\text{min}}}^1
	\frac{2\beta_1 d\beta_1}{\beta_1^2 - \beta^2}
	\br{ \Im \, Z_i(\beta_1) - \Im \, Z_i(\beta) + \Im \, Z_i(\beta)},
}
and evaluate the part with $\Im \, Z_i(\beta)$ in the numerator:
\eq{
	\Re \, Z_i(\beta)
	&=
	\frac{1}{\pi} \left\{
		\Im \, Z_i(\beta) \log\brm{\frac{1-\beta^2}{\beta_{\text{min}}^2 - \beta^2}}
		+
		\right.
		\nn\\
		&\qquad\qquad\qquad\left.+
		\int\limits_{\beta_{\text{min}}}^1 \frac{2 \beta_1 d\beta_1}{\beta_1^2 - \beta^2}
		\brs{\Im \, Z_i(\beta_1) - \Im \, Z_i(\beta)}
	\right\}.
	\label{eq.DispersionRelation}
}
The integral in braces is already regular at the point $\beta_1 = \beta$ and can be evaluated
with ease by any standard numerical method of integration.

We note that for the $D$-meson loop contribution the imaginary part $\Im \, Z_D(\beta)$ from (\ref{eq.ImZD})
is non-zero above threshold ($s > 4M_D^2$), thus the lower limit of integration in
(\ref{eq.DispersionRelation}) is $\beta_{\text{min}} = \sqrt{1 - M_p^2/M_D^2}$.
For the three gluon contribution the threshold for the imaginary part $\Im \, Z_{3g}(\beta)$
coincides with the threshold of the reaction, i.e. $s_{\text{min}} = 4M_p^2$ and thus
lower limit of integration is $\beta_{\text{min}} = 0$.


\end{document}